# Robust Sliding Mode Control for Air-to-Air Missile

Eerik Cockin, Xinhua Wang
Aerospace Engineering
University of Nottingham, UK
Email: wangxinhua04@gmail.com

**ABSTRACT**
Within the missile guidance and control system the autopilot must overcome an array of variables and uncertainties to maintain tracking trajectory. A large uncertainty explored in this paper is the difference between the assumed flight dynamics, the controller design relies on, and the true flight dynamics the missile experiences. To capture these differences experimental wind tunnel data was used to represent real life aerodynamics whereas a low fidelity wing and tube structure was used for the controller dynamics. Other variables affecting controller performance are also quantified and explored in this paper, such as the changing mass, center of gravity, dynamic pressure, actuator bandwidth and sensor noise. A second order sliding mode controller utilizing an exponential reaching law was developed to overcome the cumulative uncertainties. The designed controller is capable of a 0.2s settling time and a 3% overshoot in ideal conditions. The controller relies on measurements of both dynamic pressure and angle of attack, when a 10% and ±2° respective noise is introduced at 200Hz, the controller maintains a 5% steady state error and a time constant of 0.29s. The exponential reaching law provides superior chattering mitigation over traditional techniques like the tanh function, with no loss in controller performance.

## NOMENCLATURE

| | | |
|---|---|---|
| $\alpha$ | $rad$ | Angle of attack (AoA) |
| $e$ | - | Error |
| $m$ | $Kg$ | Mass |
| $I_{YY}$ | $kg.m^2$ | Moment of Inertia |
| $\bar{q}$ | $Pa$ | Dynamic pressure |
| $q$ | $rad\,s^{-1}$ | Pitch rate |
| $\theta$ | $rad$ | Pitch |
| $g$ | $m\,s^{-2}$ | Acceleration from gravity |
| $V$ | $m\,s^{-1}$ | Velocity |
| $L_\alpha$ | $N\,rad^{-1}$ | Lift due to angle of attack |
| $M_\alpha$ | $Nm\,rad^{-1}$ | Pitching moment due to angle of attack |
| $M_q$ | $Nms\,rad^{-1}$ | Pitching moment due to pitch rate |
| $M_\delta$ | $Nm\,rad^{-1}$ | Pitching moment due to canard deflection |
| $\delta$ | $rad$ | Canard deflection |
| $S_c$ | $m^2$ | Canard wing area |
| $S_t$ | $m^2$ | Tail wing area |
| $l_c$ | $m$ | Canard moment arm |
| $l_t$ | $m$ | Tail moment arm |
| $C_L$ | - | Coefficient of lift |
| $\Lambda_{LE}$ | $rad$ | Leading edge sweep |
| $AR$ | - | Aspect ration |
| $M$ | - | Mach number |

## 1. INTRODUCTION

The role of any missile is to intercept a target, with air-to-air missiles overcoming the additional challenge of a highly maneuverable target demanding faster and more accurate repose.

Air-to-air missiles must overcome the changing inertial conditions brought on by the burn and expulsion of solid propellant alongside the Mach varying aerodynamic magnitudes and locations. Specifically, the Mach varying aerodynamic center of lift is very difficult to estimate without the use costly CFD and supersonic wind tunnel testing. Even with experimental data, transonic performance can vary dependent on unpredictable atmospheric conditions. In this paper a robust controller is designed to overcome these uncertainties whilst maintaining a performance comparable to other current controllers. Namely a time constant $< 0.35\,s$ and a following error $< 5\%$.

## 2. BACKROUND
### 2.1 Guidance System

The guidance loop uses Line of Sight (*LOS*) data to generate autopilot command values. The command values are then modulated by the autopilot using data from the body sensors, as shown in Figure 1.

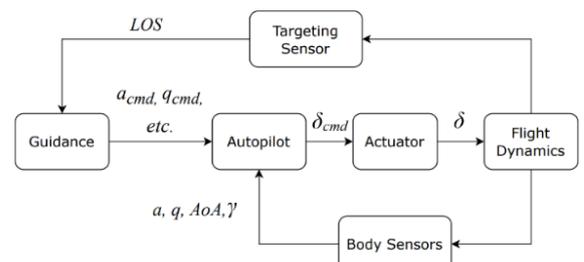

**Figure 1: Missile guidance configuration.**

Many different autopilot command structures can be used dependent on the sensor information available, the most common being acceleration ($a$), pitch rate ($q$), Angle of Attack (AoA) or flight path angle ($\gamma$). While only one command value is necessary most missiles will use different autopilot command structures dependent on the stage of the mission [1].

Surface to air missiles typically use a flight path command for launch whereas AoA commands are used for agile turns as rapid control of the missile's velocity vector is required. Acceleration autopilots are generally preferred for midcourse and terminal phases [2]. Alongside performance, other factors like packaging constraints and cost must be considered in a trade-off analysis. For example, an IMU measures acceleration and pitch rates to a very high accuracy however the cost is equally high for a one time use missile.

## 2.2 Autopilot

As shown in Figure 1 the commanded value is then controlled to make the error between the desired value and the measured value, from the sensors, equal to zero; $e \equiv (\alpha_{cmd} - \alpha) \to 0$. This controller section is the main problem explored in this paper.

The most common approach to missile control is to split the flight envelope into different bands and linearize the flight dynamics about one point so that the linear control is acceptable within each band, known as Linear Time Invariant (LTI) controllers. The controllers' properties then switch from band to band to provide acceptable control over the entire flight envelope. An example of this is shown by Siouris [3] for bands split over variable Mach numbers and altitudes.

Gain scheduling is the most comon form of missile control [4] and easy to impliment with the LTI system. Predetermined gains are stored in the form of lookup tables and implemented depedant on operating conditions.

LTI controllers are simple to design and can produce exceptional results as the controller is optimised at each design point. However, to accurately linearize flight dynamics at a certain design point a high accuracy of aerodynamic performance is required. Also, the scheduled gains are not smooth, this can reduce controller performance in regions between LTI design points.

The negatives of the LTI controllers can be avoided throught the use of a Robust controller. A robust controller is capable of maintaining stability over the entire flight envelope and capable of rejecting distubances without the use of pre-scheduled switching criteria. Becasuse a single set of non optimised cotroller parameters are used response is typically lost in the process.

The simplest form of robust controller is to increase the loop gain, as the loop gain becomes infinte the error of the system vanishes however this is only operable in theory as factors such as controller saturation, noise, and dynamic instability cause major issues. [4]

Robust forms of gain scheduling are heavily researched topics with many forms of adaptive scheduling developments. [5] shows an automated method for tuning gain magnitudes for optimal control at varying flight conditions. Such a method negates the unoptimized nature of robust control whilst avoiding the time-consuming need for large look-up tables. [6] uses a separate method of adaptive pole placement to select gains to cover different, known, aerodynamic conditions. The controller shown does maintain robustness at each design point to overcome the unknown mass variation.

Another popular controller is the $H_\infty$ loop shaping method. The main benefit of using a loop shaping controller is that it is smooth, unlike with gain scheduling where the controller parameters jump from band to band, the loop shaping gradually bridges the gap between two operating conditions. This smoothness, however, typically comes with a reduction in response time [7], [8].

The final commonly used controller is the Sliding Mode Controller (SMC), it operates off the basic philosophy that the fastest way to a desired value is to use maximal control input in the direction of interest [4].

[9] shows a robust higher order SMC capable of negating significant disturbances to the assumed flight model whilst maintaining stability of the missile through unstable phases.

A common issue faced by the SMC is a phenomenon known as chattering. The dramatic enforcement of control input causes rapid switching of the control surface, which decreases performance and is harmful to actuators, accelerating the effects of wear [10].

The performance of an autopilot is typically analysed using 3 metrics: time constant, steady state following error and overshoot. A time constant of 0.35s is the maximum allowable time [8] however newer controllers have increased the standard to 0.2s [11] with a relatively standard steady state error of 5%. Overshoot is sometimes not measured in controller analysis; this can lead to significant performance implications as a large overshoot leads to longer settle times [7]. Maximum allowable overshoots as low as 3% can be observed in advanced controllers [11].

All adaptive gain selection techniques require very accurate approximations for flight dynamics at selected design points however, this isn't always possible given the need for expensive supersonic testing. A robust sliding mode AoA controller is therefore explored in this paper to overcome the uncertainties related to flight dynamics.

## 2.3 Missile Dynamics
### 2.3.1 Truth vs Design model

A concept explored heavily by Friedland [4] for control system simulation is a distinction between the "Truth model" and the "Design model". Where the truth model mimics real physics as accurately as possible whereas the design model uses simplified equations, often linearized, which can then be solved in the context of control. This distinction seems practical in aerospace systems given their highly coupled non-linear dynamics, full of idealisations.

### 2.3.2 Flight Dynamics Models

The distinction between truth and design models has not been adopted for missile control simulations. Many papers use identical models for both the controller and the plant, or use the same flight model to perform adaptive controller selection and then test said controller using the same flight model [6], [12].

The level of flight model fidelity varies dramatically across different papers ranging from a simple 1 Degree of Freedom (DoF) model [11] up to non-linear 6 DoF models [2], requiring vast aerodynamic data. Furthermore, different levels of component fidelity are adopted, actuators being the most important as it is the main component that limits the bandwidth of the missile [13].

This compounded assortment of varying flight model conditions makes analysis of the control system problematic for its inherent link to flight model fidelity.

## 2.4 Missile Specific Stability

Missiles have a wide variety of different factors that affect controllability, many can be measured or ignored, however some have unpredictable consequences.

### 2.4.1 Dynamic Pressure

Dynamic pressure is defined by $\bar{q} = \frac{1}{2}\rho V^2$. Air density, $\rho$, has very little change over the flight envelope however velocity, $V$, changes significantly and, given its second order, has a very large impact on performance. Given this significance velocity is often directly or indirectly measured.

### 2.4.2 Inertial Properties

The missile produces thrust by burning the solid propellant aft of the missile, this shifts the centre of gravity forward hence changing the moment arm and effectiveness of control surfaces. The mass and moment of inertia will also decrease resulting in increased mobility of the missile. While the burn sequence is rapid and has a significant effect on stability the burn profile is relatively predictable. Given the grain type and packing, performance can be calculated and tested easily.

### 2.4.3 Aerodynamic Properties

Both the magnitude of the coefficient of lift and the center of lift locations vary with Mach numbers, particularly in transonic regions. These variations require expensive and time consuming CFD and supersonic wind tunnel experiments. While the impact of Mach variations is large, they change on a slow time scale making control more manageable.

# 3. TRUTH MODEL
## 3.1 Assumptions

The high-fidelity truth model was derived based on the following assumptions:
- A1. The air frame is rigid and any aeroelastic effects are neglected.
- A2. The rotational moment of inertia ($I_{yy}$) is constant.
- A3. Lift coefficient is linear in between positive and negative stall angles.
- A4. Earth is locally flat; hence the curvature and rotation of the Earth can be ignored.

A1. and A4. are standard modelling assumptions for sub-hypersonic flight $M < 5$. A2. was assumed as it greatly simplifies the flight model and bares little effect on stability. A non-decreasing $I_{yy}$ will however increase time to reach a desired pitch angle.

A3. has the largest impact as [14] shows that this linearity is not the case for most Mach numbers however, implementing this data set would require massive look up tables severely effecting simulation run time making controller analysis highly impractical.

## 3.2 Degree of Freedom

The 6 DoF model is standard for aircraft modelling, however the configuration of missiles allows for some simplification. The pitch and yaw axis are identical, so the yawing moments and the side force ($r, Y$) were neglected.

In modern missiles Skid to Turn (STT) manoeuvres are most widely adopted [13], the pitch and yaw meet desired AoA and sideslip angles to meet a desired trajectory. These are preferred over traditional Bank to Turn (BTT) manoeuvres, where a bank angle is first met then the pitch meets the desired trajectory. STT is faster than the BTT and due to the relatively low inertial cross coupling between roll, pitch, and yaw [3] the roll dynamics ($p$) do not need to be modelled.

After simplifications the standard longitudinal 3 DoF equations remain shown in Eq.*(1)* where the forces $F_{x,z}^B$ are resolved in the body axis.

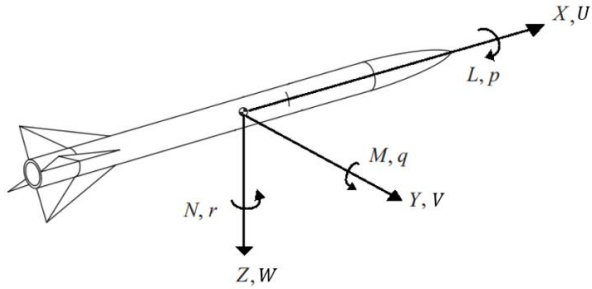

Figure 2: 6 DoF notation

$$\begin{bmatrix} \dot{u} \\ \dot{w} \\ \dot{q} \end{bmatrix} = \begin{bmatrix} -qw + F_x^B/m \\ qu + F_z^B/m \\ M_y/I_{YY} \end{bmatrix} \quad (1)$$

## 3.3 Thrust Profile

Thrust is produced by the burning of the solid propellant, with the magnitude of the thrust being proportional to the area of propellant being burnt at any given time. This means the time varying thrust profile can be selected to meet the mission requirements by packing the propellant in a certain orientation. Short range, highly maneuverable missiles use an all-boost profile where all the propellant is burnt quickly.

The all-boost burn sequence follows a semi-sinusoidal profile shown in Figure 3.

## 3.4 Change of Mass

The mass of the missile reduces as the solid propellant is burnt, a numerical approximation for this relation was developed by [3] and adapted to form Eq. *(2)*.

A fuel mass ($m_f$) equal to 40% of the initial mass ($m_0$) was selected in line with [6].

Although all the aerodynamic data was provided by NASA [14], there was no mention of mass or moment of inertia. An estimate for $m_0$ was taken by scaling down the mass of current air-to-air missiles maintaining a similar wing loading. The moment of inertia was then scaled down proportional $\propto \Delta m \cdot \Delta l^2$ producing a final value of $m_0 = 10\ kg$ and $I_{YY} = 1\ \text{kg.m}^2$

The final values of mass and moment of inertia were validated by observing the short period pitch oscillation response of the finalised truth model.

$$m = m_0 - m_f \frac{\int_0^t T(t)}{\int_0^\infty T(t)} \quad (2)$$

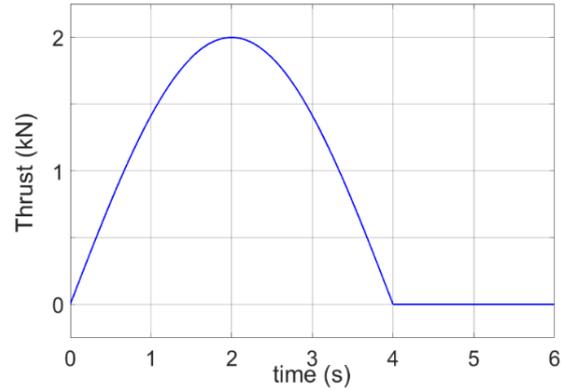

Figure 3: Thrust profile over boost phase

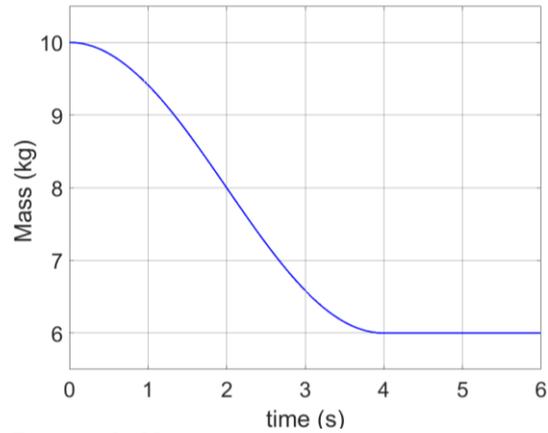

Figure 4: Missile mass over boost phase

## 3.5 Change of Center of Gravity

The missile was split into two subsections, a static and a dynamic part representing the missile body and propellant respectively.

The static mass of the missile body ($m_b$) was given an individual Center of Gravity (CoG) at 35% along the missile length and the initial mass of the propellant was given a CoG at 80% along the missile length. While the mass of the propellant changes it is assumed that its individual CoG does not, as the propellant is burnt inside to out.

These values were chosen as they create a reasonable initial and final CoG as depicted in Figure 5.

$$CG = \frac{0.35\ l \cdot m_b + 0.8\ l\ (m(t) - m_b)}{m(t)} \quad (3)$$

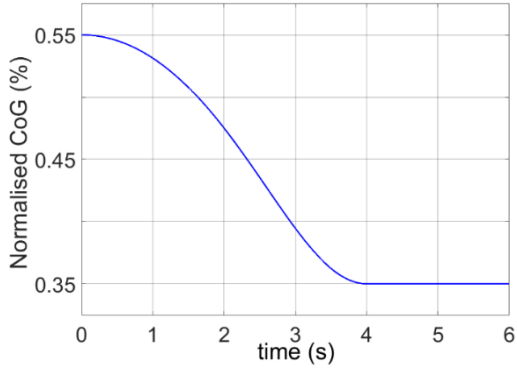

**Figure 5: Normalised Centre of Gravity over boost phase**

### 3.6 Atmospheric Model

The atmospheric conditions vary relatively predictably as a function of altitude ($h$). The standard NASA atmospheric model was used [15] shown in Eqs. *(4)* and *(5)*. Using the pressures ($P$) and temperatures ($\mathbb{T}$) at each altitude the air density and speed of sound could be calculated.

For the troposphere:

$$\mathbb{T} = 252.14 - 0.00649h$$
$$P = 101.29 * \left(\frac{\mathbb{T}}{288.08}\right)^{5.256} \quad (4)$$

For the lower stratosphere:

$$\mathbb{T} = 216.64$$
$$P = 22.65 * e^{1.73 - 0.000157h} \quad (5)$$

### 3.7 Aerodynamic Model

The Mach varying aerodynamics are a key part of the control problem so using realistic magnitudes was important. NASA wind tunnel data [14] was collected and formatted into look up tables to capture the Mach varying aerodynamic properties as shown in blue by Figure 6. Note the coefficient of drag was also used in the truth model however bares no relevance to pitch control so not shown.

### 3.8 Actuator

The actuator is the main component that limits the bandwidth of the missile autopilot [13]. A 2nd order filter was used to represent this bandwidth selection.

$$\frac{\delta}{\delta_c} = \frac{\omega_n^2}{s^2 + 2\xi\omega_n + \omega_n^2} \quad (6)$$

Typically, papers use an angular frequency of $\omega_n \sim 150 - 200 \; rad/s$ [12] [16] for actuators however, the control surfaces in this small missile experience roughly 10 times lower loads so a conservative increase to $500 \; rad/s$ was decided with a damping factor $\xi = 0.7$.

A rate limiter was also introduced so that maximum angular velocity achievable is $\omega_n$.

The maximum deflection of the actuator was limited using a saturation function of 0.3 rads (~17°). Ideally the saturation would be AoA dependent so the total incidence AoA into the canards would be less than stall angle however this is more related to the missile configuration trade off hence not necessary for control analysis.

## 4. DESIGN MODEL
### 4.1 Equations of Motion

The axial force equation bears no relation to pitch dynamics hence it was neglected leaving the pair of coupled differential equations from Eqs. *(7)* and *(8)* representing the nominal forces and pitching moments respectively.

$$\dot{\alpha} = q - cos(\theta)\frac{g}{V} + \frac{L_\alpha}{mV}\alpha \quad (7)$$

$$I_{YY} \cdot \dot{q} = M_\alpha \alpha + M_q q + M_\delta \delta \quad (8)$$

Eqs. *(7)* and *(8)* were combined to make Eq.*(9)*.

$$\ddot{\alpha} = \frac{M_\alpha}{I_{YY}} \cdot \alpha + \frac{M_\delta}{I_{YY}} \cdot \delta + d$$

$$where: \quad d = \Delta\frac{M_\alpha}{I_{YY}} \cdot \alpha + \Delta\frac{M_\delta}{I_{YY}} \cdot \delta + \frac{M_q}{I_{YY}}q \quad (9)$$

$$- q \cdot sin(\theta)\frac{g}{V} + \frac{L_\alpha}{mV}\dot{\alpha}$$

$M_\alpha$ & $M_\delta$ are large relative to the other terms therefore the other terms are rejected as disturbance. The smaller terms could be calculated given measurement instruments were introduced for $q$ and $\dot{\alpha}$ however the performance benefit would not be worth the added complexity and weight.

### 4.2 Aerodynamic Derivatives

As mentioned in section 2.4.3, the Mach varying center of lift is very unpredictable, a static location was chosen at the ¼ chord of the aerofoils for the moment to act through. These simplistic assumptions produce Eqs. *(10)* & *(11)*.

$$M_\alpha = \bar{q}C_L(S_c l_c - S_t l_t) \quad (10)$$

$$M_\delta = \bar{q}C_L S_c l_c \quad (11)$$

$S_{c,t}$ are known from geometry provided by NASA [14], $\bar{q}$ is measurable, $l_{c,t}$ and $C_L$ are estimated using crude assumptions. The effectiveness of these assumptions can be seen relative to the truth model data in Figure 6.

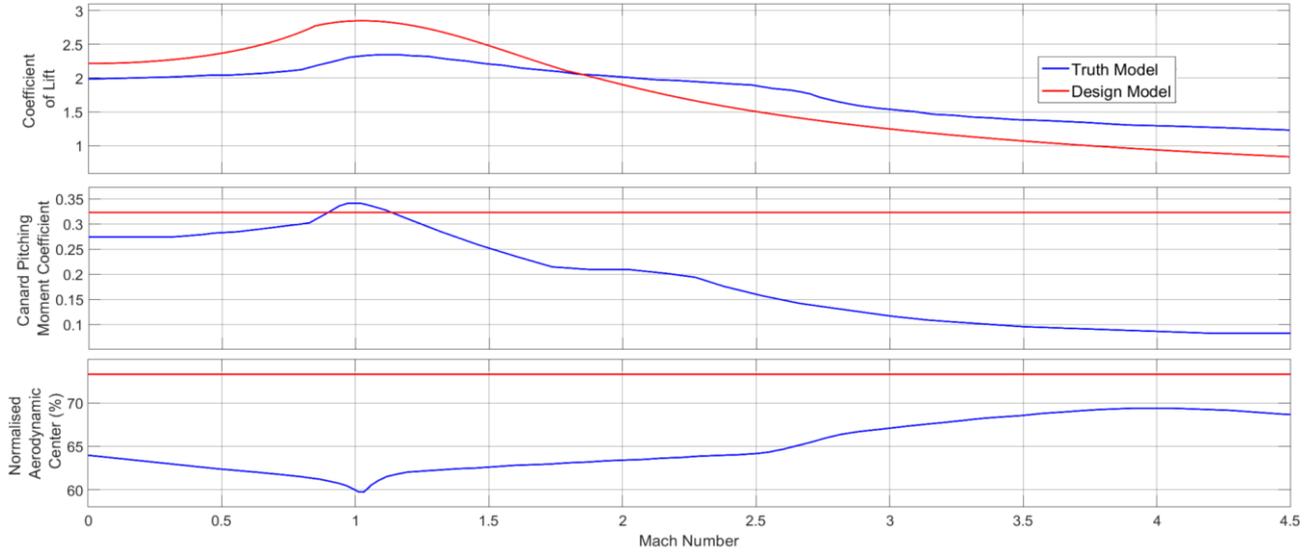

Figure 6: Aerodynamic coefficients used in Design and Truth models

## 4.3 Lift Coefficient

All derivations are based off symmetric thin aerofoil assumptions on a tube and wing structure, ignoring any lift provided by the body. The Mach varying lift co-efficient was estimated using empirical approximations from Houghton & Carpenter [17] and Cook [18].

Sub sonic approximations are well understood and long standing with many of Prantl's rules still holding true. Eq.*(12)* shows the incompressible lift coefficient of an infinite span aerofoil ($C_L^{2D}$) given the thickness to chord ratio ($t/c$) and leading-edge sweep, Eq.*(13)* then corrects for compressibility and induced tip vortex losses.

$$C_L^{2D} = 1.8\pi \left(1 + \frac{t}{c}\right) \cos(\Lambda_{LE}) \quad (12)$$

$$C_L = \frac{1}{\sqrt{1 - M^2 \cos^2(\Lambda_{LE})}} \cdot \frac{C_L^{2D}}{1 + \frac{C_L^{2D}}{\pi \cdot AR}} \quad (13)$$

Supersonic assumptions are more complex as they are dependent on shock location hence require information about the wing and aerofoil geometry. A simplified assumption was made using Eqs. *(14)* & *(15)* where $M > sec(\Lambda_{LE})$.

$$C_L^{2D} = \frac{4\cos(\Lambda_{LE})}{\sqrt{M^2 \cos^2(\Lambda_{LE})}} \quad (14)$$

$$C_L = C_L^{2D} \left(1 - \frac{1}{2AR\sqrt{M^2 \cos^2(\Lambda_{LE}) - 1}}\right) \quad (15)$$

No solid transonic approximations exist so, for the region $0.85 < M < sec(\Lambda_{LE})$ the supersonic and subsonic regimes were visually bridged using points connected and smoothed by a high order polynomial as shown in Figure 6.

## 5. SLIDING MODE CONTROLLER
### 5.1 Controller Design

The bulk of controller design follows work from Liu and Wang [10].
The design model AoA equation is second order, Eq.*(9)*, hence a second order sliding function was selected, Eq.*(16)*.

$$s = \dot{e} + ce \quad (16)$$

The controller, Eq.*(17)*, was derived from the design model equation of motion, Eq. *(9)*, and $v$ is the auxiliary controller described by Eq. *(18)*.

$$\delta = I_{YY}\left(\frac{v - \frac{M_\alpha}{I_{YY}}\alpha}{M_\delta}\right) \quad (17)$$

$$v = c\dot{e} + \ddot{\alpha}_{cmd} - \eta\, sgn(s) \quad (18)$$

The Lyapunov function is defined by Eq. *(19)*. As the function approaches zero the system becomes stable. Stability can therefore be forced by making the derivative of the Lyapunov function negative.

$$V = \frac{1}{2}s^2 \quad (19)$$

$$\dot{V} = s\dot{s} = s(c\dot{e} + \ddot{\alpha}_d - \ddot{\alpha}) \quad (20)$$

By substituting Eq. *(9)* into Eq. *(20)* then combining with Eqs. *(17)* and *(18)* the stability condition from Eq. *(23)* was derived. This shows that stability can be maintained where $\eta$ is larger than the cumulative disturbance.

$$\dot{V} = s(d - \eta \cdot sgn(s)) = ds - \eta|s| \quad (21)$$

$$d_{max} - \eta \leq 0 \quad (22)$$

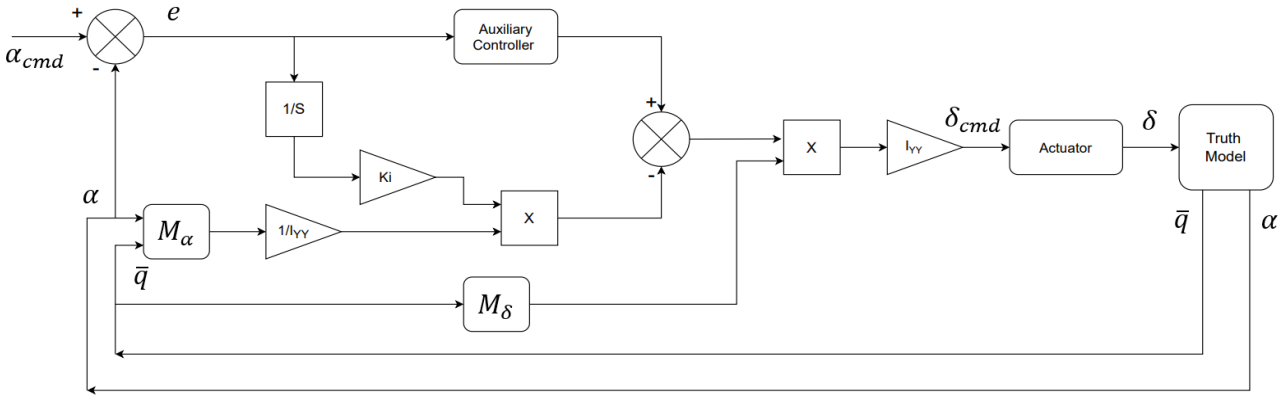

**Figure 7: Control system simulation overview**

## 5.2 Reaching Law and Chattering

Within the auxiliary controller, Eq. *(18)*, exists the reaching law, $sgn(s)$, its purpose is to bind the function to the sliding surface by overcoming disturbance as shown in Eq. *(21)*. However, the issue of using a $sgn(s)$ reaching law is that it is discontinues about $s = 0$, therefore when the system is near stable the commanded canard deflection rapidly oscillates from maximum to minimum.

Chattering also occurs when the bandwidth of a system is too wide. The $sgn(s)$ function causes strong control inputs even when the system is near stable so a suitably narrow bandwidth is required to catch any over corrections. Bandwidth needs to be narrower for a lighter, more agile missile.

Common remedies for chattering include the use of a $tanh(s)$ function instead of the $sgn(s)$ function or a power reaching law [10] shown in Eq. *(23)*.

$$u_r = \eta \cdot |s|^a \cdot sgn(s)$$
where: $\quad 0 < a < 1 \quad$ (23)

Both options remove the discontinuity in the reaching law for a smooth control input.

## 6. RESULTS
### 6.1 Estimation of Aerodynamic Uncertainties

Figure 7 shows the structure of the controller and what sensor data is required for each function. The controller was slightly augmented from Eq. *(17)* to include an integral term as an observer into the $M_\alpha$ loop. The slow time varying nature of the centre of lift uncertainty means the simple observer provided acceptable performance.

However as seen in, Figure 8 and Figure 9, performance is compromised from 3 to 5 seconds; this is right at the end of the boost phase so maximum velocity is achieved and the control surface effectiveness is at its minimum. For a more robust system a higher order observer could be used to cope with the non-linear variations, however this is outside the scope of this project.

### 6.2 Chattering Mitigation
#### 6.2.1 First Order SMC

Figure 8 and Figure 9 show that chattering persists despite the use of a continues reaching law, with the canard deflection rapidly fluctuating from maximum to minimum.

Given a continuous reaching law was used in both simulations, the remaining chattering is a consequence of the limited bandwidth and subsequent overcorrections enforced by the high gain controller.

Figure 9 shows the superior chattering mitigation characteristics of the $tanh$ function over the power reaching law shown in Figure 8, specifically during the early stages of flight. Using Figure 10 the difference in reaching law performance can be analysed. While the two functions appear to have a similar profile, the relative difference in outputs becomes significant as $s \to 0$. By taking the first derivative of each function the difference becomes clear. For the power reaching law: $du/ds = \infty$ at $s = 0$ whereas $du/ds = 1$ at $s = 0$ for the $tanh$ function. Furthermore, the derivative of $tanh$ is smooth whereas the derivative of the power reaching law has an asymptotic profile about $s = 0$. This combination of a lower output and a lower rate of change of output gives the $tanh$ function its chattering mitigation properties. Figure 9 also shows the $tanh$ functions limitations; as the boost phase ends and the canard effectiveness decreases the controller struggles to maintain the high gain robustness required. This is a consequence of the saturation of the $tanh$ function with outputs approaching $\pm 1$.

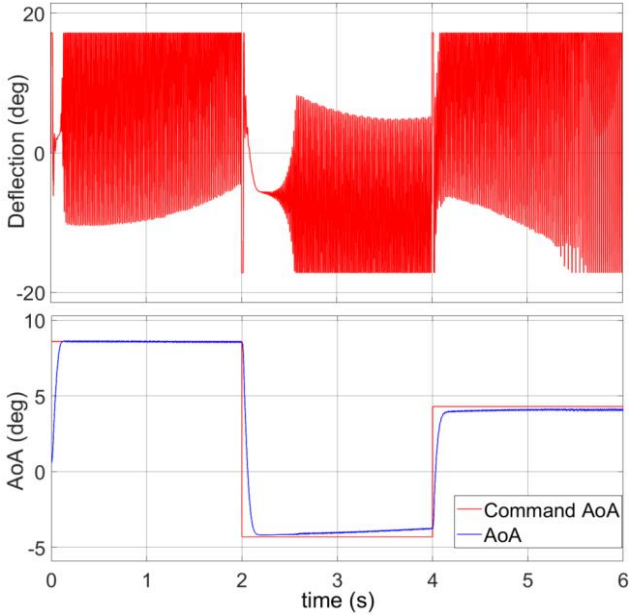

**Figure 8: Power reaching law performance**

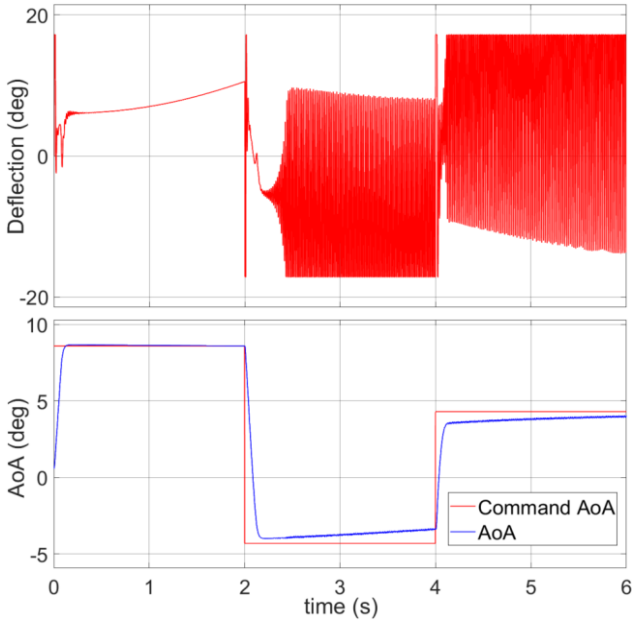

**Figure 9: tanh reaching law performance**

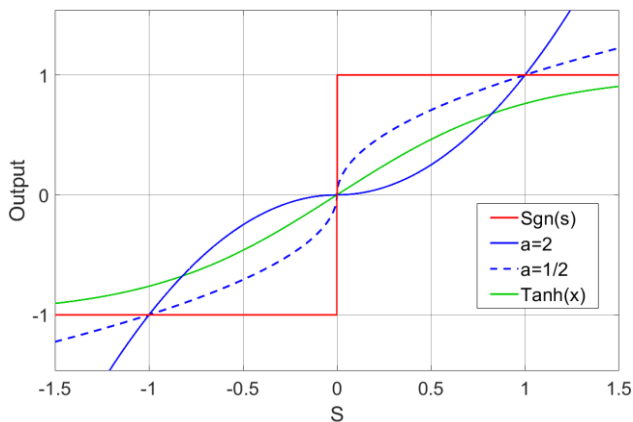

**Figure 10: Reaching law alternatives**

### 6.2.2 Second Order SMC

Further chattering mitigation was achieved through the use of a higher order SMC. A super twisting SMC [19] was then introduced to the auxiliary controller. The super twisting SMC, Eq. *(24)*, was selected over the standard twisting SMC as no real time measurements for $\dot{s}$ are required.

$$u_r = \eta \cdot |s|^a \cdot sgn(s) + b$$
$$\dot{b} = \eta_1 sgn(s) \quad (24)$$

where: $0 < a < 1$

The increased effectiveness of the controller meant a lower value of $\eta$ could be taken to achieve satisfactory results, which in turn reduced chattering, but not removed, as large oscillations remained.

Using the conclusions drawn from section 6.2.1 a reaching law smooth to the first derivative and a small $du/ds$ at $s = 0$ was required. To meet this, Eq.*(24)* was adapted to use an exponential reaching law ($a > 1$)

This is not typically used as, mathematically, the system cannot reach $s = 0$ within a finite time because $du/ds \to 0$ as $s \to 0$.

While the system is not mathematically stable the exponential nature of the reaching function combined with the high gain means $s \to 0$ withing a very short time. And the region where control is weak is far narrower than the target 5% steady state error.

The implementation of a higher order exponent in the reaching law provides substantial chattering dampening without sacrificing any loss in performance.

Performance metrics were taken from Chen et al [11] for its ideal, no noise, analysis. A maximum overshoot of 3% was set with a target settling time of 0.2s. Both metrics were achieved for the first and second step inputs, as seen in Figure 11, however given the observer shortcomings mentioned earlier, a slightly longer settling time of 0.3s was achieved for the final step input.

Chen [11] assumes a zero-disturbance model, whereas Figure 11 achieves the given performance under the substantial disturbance brought on by the truth and design model split.

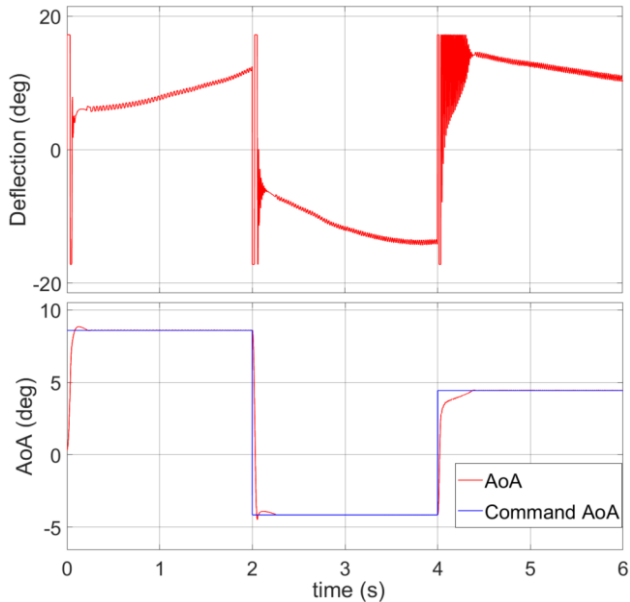

**Figure 11: Super Twisting SMC with exponential reaching law**

## 6.3 Noise Analysis

A pitfall of the sliding mode controller is its inability to handle noise, the traditional $sgn(s)$ reaching law amplifies perturbations causing further chattering.

Noise was introduced into the system for both the measured values required for control, AoA and dynamic pressure. The AoA was given a random noise magnitude of $\pm 2°$ and a velocity error of $\pm 10\%$, both at a frequency of $200\ Hz$.

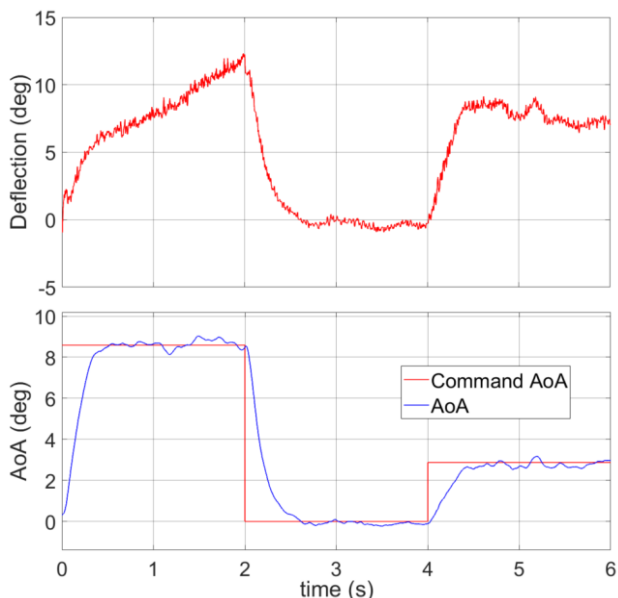

**Figure 12: Noise mitigation performance**

A simple 2$^{nd}$ order filter was used to reject the noise form the sensors, as shown in Eq. *(6)*. A cut-off frequency of $100\ Hz$ and damping ratio $\xi = 1.0$ was used.

As shown in Figure 12 the system is capable of rejecting noise and maintaining a $5\%$ steady state following error with a time constant $0.29\ s$.

## 7. FURTHER WORK

In this paper the controller coefficient's, $\eta_1, \eta, c\ \&\ K_i$ were all selected via trail and error over a handful of iterations. Further work is recommended in optimising the method of selecting these gains for the exponential high order SMC.

The effects of actuator bandwidth were heavily explored, however the separate issue of time lag was not. Further research should be conducted on the exponential reaching laws ability to overcome a time lag, through the use a phase lead compensator or other predictive modeling techniques.

## 8. CONCLUSION

This paper presents a robust angle of attack controller capable of overcoming the uncertainties associated with air-to-air missile control. The derivation of a high-fidelity truth model is shown capturing the niche variations and control limitations that occur across the flight envelope, such as the moving centre of gravity and actuator bandwidth. A low fidelity, wing and tube, flight model was developed for use in the angle of attack controller.

A robust second order sliding mode controller featuring an exponential reaching law was used. The exponential reaching law shows significant improvement in chattering mitigation over traditional methods, such as tanh functions. The uncertainty of the Mach varying aerodynamic coefficients is overcome whilst maintaining low settling times, overshoots and steady states errors.

Such a robust controller shows the feasibility of avoiding costly supersonic wind tunnel experiments potentially reducing overall missile design cost.